\documentclass[12pt]{iopart}

\newcommand{\mbr}{{\mathbf{r}}}

\newcommand{\hr}{{\hat{\bf r}}}

\newcommand{\cP}{{\cal P}}

\newcommand{\im}{{\mbox{i}}}
\newcommand{\la}{\label}

\newcommand{\beq}{\begin{equation}}
\newcommand{\eeq}{\end{equation}}
\newcommand{\bay}{\begin{eqnarray}}
\newcommand{\eay}{\end{eqnarray}}
\begin{document}
\title[Closed form representation for a projection onto Coulomb bound states]
{Closed form representation for a projection onto infinitely
dimensional subspace spanned by Coulomb bound states}
\author{O.M. Deryuzhkova$^1$, S.B. Levin$^{2,3}$, S.L. Yakovlev$^{2,4}$ }
\address{$^1$Department of Physics, Gomel State University, Gomel,
Belarus 246019}
\address{$^2$Department of Computational Physics, St Petersburg
University, St Petersburg, Russia 198504 }
\address{$^3$Department of Physics, Stockholm University, Stockholm,
Sweden 10691}
\address{$^4$Department of Physics and Astronomy,
California State University at Long Beach, CA 90840}

 \ead{yakovlev@cph10.phys.spbu.ru}
 \ead{levin@physto.se}
%
\begin{abstract}
The closed form integral representation for the projection onto the
subspace spanned by bound states of the two-body Coulomb Hamiltonian
is obtained. The projection operator onto the $n^2$ dimensional
subspace corresponding to the $n$-th eigenvalue in the Coulomb
discrete spectrum is also represented  as the combination of Laguerre
polynomials of $n$-th and $(n-1)$-th order. The latter allows  us to
derive an analog of the Christoffel-Darboux summation formula for
the Laguerre polynomials. The representations obtained are believed to be
helpful in solving the breakup problem in a system of three charged
particles where the correct treatment of infinitely many bound
states in two body subsystems is one of  the most difficult
technical problems.
\end{abstract}
\submitto{\jpb} 
\maketitle
\section{Introduction}
The two-body Coulomb problem is perhaps the most famous problem of
quantum mechanics formulated on the basis of the  Schr\"odinger wave
equation \cite{schrd}. First solved analytically for the bound
states \cite{schrd}, it was then solved  for scattering states
\cite{YWB}, as well as for the Green's function \cite{WW},
\cite{hostler}, \cite{Schwinger}. The situation is quite different
for the systems involving three and more charged particles. There is
no analytic solution to the Schr\"odinger equation in this case and
the problem exhibits a great complexity especially if the ionization
process is energetically permitted. In this case, the infinitely many
open excitation channels are lead to  extremely complicated
behavior of the wave function, which asymptotically possesses infinitely
many terms. Although  definite progress
has been made
in the practical
numerical solution of the three charged particles problem above the
disintegration threshold 
\cite{RescignoBray} by methods avoiding the explicit use of the wave
function asymptotics, the theoretical status of the few-body Coulomb
problem is still  unsatisfactory in many respects. Among the very
extensive literature devoted to the few-body Coulomb problem,  the
following works  \cite{Rudge}, \cite{Peterkop} and
\cite{FaddeevMerkuriev}  refer to the theoretical aspects
of the problem.

Recently we have presented a new method of handling the Coulomb
potentials in the few-body Hamiltonian with the help of the
Coulomb-Fourier  transform (CFT) \cite{AltLevinYakovlev}. The method
allows us to exclude the long-range Coulomb interaction from the
Hamiltonian by a specially constructed unitary transformation. This
method was proven  useful for repulsive Coulomb interactions. In
the case of attraction, the analytic closed form representation for
the projections onto the Coulomb bound-states subspace may lead to a
substantial simplification of the CFT method machinery. Although the
latter was the primary goal,  it was found  that the
representations obtained, being quite general, are not well addressed
in the literature, and this has  stimulated this publication.

The paper is organized from three sections. After the introduction, in the section two we
derive representations for
projection operators and consider some particular cases.
The third section concludes the paper.
Throughout the paper the bold letters, e.g.  $\mbr,\mbr'$, are used for vectors and not bold for their
magnitudes, e.g. $r=|\mbr|$. The unit vector associated with $\mbr$ will
be denoted by $\hr=\mbr/r$.
\section{Representations for projections}
We consider the infinite dimensional projection operator
\beq
\cP_d=\sum_{n=1}^{\infty}\cP_n, \ \ \cP_{n_{1}}\cP_{n_{2}}=\delta_{n_1 n_2}\cP_{n_1}.
\la{P_d}
\eeq
The operators $\cP_n$
are the orthogonal projections onto the $n^2$-dimensional subspaces spanned by the two-body Coulomb
bound-states $\langle \mbr|\psi_{nlm}\rangle=\psi_{nlm}(\mbr)$. $\cP_{n}$
have kernels
\beq
\cP_n(\mbr,\mbr')=\sum_{l=0}^{n-1}\sum_{m=-l}^{l}\psi_{nlm}(\mbr)\psi^{*}_{nlm}(\mbr'),
\la{P_n}
\eeq
where the normalized  Coulomb bound-state wave functions are chosen in the form
\beq
\psi_{nlm}(\mbr)= \alpha^{3/2}\frac{2}{n^2}\sqrt{\frac{(n-l-1)!}{(n+l)!}}
\left(\frac{2\alpha r}{n}\right)^{l}e^{-\frac{\alpha r}{n}}
L^{(2l+1)}_{n-l-1}\left(\frac{2\alpha r}{n}\right)Y^{m}_{l}({\hr}).
\label{psi}
\eeq
Here $\alpha= \frac{\mu e^2}{\hbar^2}Z_1Z_2>0$, functions
$L^{(k)}_{n}$ and $Y^m_l$ are the generalized Laguerre polynomials and spherical harmonics as
they are defined in \cite{abram},
respectively.
The wave-function $\psi_{nlm}(\mbr)$ obeys the Schr\"odinger equation 
\beq
(H-E_n)\psi_{nml}(\mbr)=(-\Delta_{\mbr} -\frac{2\alpha}{r}-E_n)\psi_{nml}(\mbr)
=0
\la{Schr-e}
\eeq
with $E_n=-\alpha^2/n^2$ and $n$ positive integer.

In order to work out the representations for $\cP_{d,n}$ we are seeking, let us begin with
the standard
formula for the  projection $\cP_n$ as the residue of the Green's function
\beq
\cP_{n}(\mbr,\mbr') = \frac{1}{2\pi \im}\oint_{C_{E_n}}{\cal G}_c(\mbr,\mbr',\zeta)d\zeta
\la{P_ci}
\eeq
with the contour $C_{E_n}$ encircling the point $E_n$ in positive direction in the $\zeta$ complex plane.
Then, using the Hostler \cite{hostler} 
representation for ${\cal G}_c(\mbr,\mbr',\zeta)$
\beq
{\cal G}_c(\mbr,\mbr',\zeta)=\frac{\Gamma(1-\im\nu)}{4\pi|\mbr-\mbr'|}\frac{1}{\im \sqrt{\zeta}}
\left(\frac{\partial}{\partial s_{+}}-\frac{\partial}{\partial s_{-}}\right)
W_{\im \nu,\frac12}(-\im\sqrt{\zeta}s_{+})M_{\im \nu,\frac12}(-\im\sqrt{\zeta}s_{-})
\label{host}
\eeq
where $\nu=\alpha/\sqrt{\zeta}$ and $s_{\pm}=r+r'\pm|\mbr-\mbr'|$ and evaluating the residue,
 we arrive at the following expression for
$\cP_n(\mbr,\mbr')$
\beq
\cP_n(\mbr,\mbr')=\frac{\alpha^3}{n^4}\frac{e^{-\frac{\alpha}{2n}(s_{+}+s_{-})}}{\pi(s_+-s_-)}
\label{pro1}
\eeq
$$
\times\left[s_+L_{n-1}\left(\frac{\alpha
s_-}{n}\right)L_{n-1}^{(1)}\left(\frac{\alpha s_+}{n}\right)-
s_-L_{n-1}^{(1)}\left(\frac{\alpha
s_-}{n}\right)L_{n-1}\left(\frac{\alpha s_+}{n}\right)\right].
$$
Alternatively, we can transform (\ref{pro1}) into the form
\beq \cP_n(\mbr,\mbr')=\frac{\alpha^2}{n^2}
\frac{e^{-\frac{\alpha}{2n}(s_++s_-)}}{\pi(s_+-s_-)}
\label{proj13}
\eeq
\nopagebreak
$$
\times\left\{L_{n-1}\left(\frac{\alpha
s_+}{n}\right)L_n\left(\frac{\alpha s_-}{n}\right)-
L_n\left(\frac{\alpha s_+}{n}\right)L_{n-1}\left(\frac{\alpha
s_-}{n}\right)\right\}.
$$
The formulae (\ref{pro1}) and (\ref{proj13}) are the starting points for representations of this paper.

\subsection{Integral representation for  projections onto the discrete spectrum subspace}
To proceed with formulae (\ref{pro1}) and (\ref{proj13}), we use the well known expression for Laguerre
polynomials \cite{Bateman} in terms of Bessel functions
\beq
L_{n-1}^{(\beta)}\left(\frac{y}{n}\right)=\frac{n^n}{n!}n^{\beta+1}y^{-\beta/2}
e^{y/n} \int_0^\infty dx \,
x^{\beta/2-1}\left(xe^{-x}\right)^n
J_\beta\left(2\sqrt{xy}\right)
\label{Laguerre}
\eeq
and the following integral representations (3.382.7) \cite{Grad}
$$
\frac{n^n}{n!}=\frac{1}{2\pi}e^n\int_{-\infty}^\infty
\frac{dy}{(1-\im y)^n}e^{-\im ny}
$$
and (6.631.10) \cite{Grad}
$$
\frac{1}{n}e^{\alpha(a+b)/2n}=\int_0^\infty dy\,
e^{-ny}I_0\left(\sqrt{2y\alpha(a+b)}\right)
$$
where $J_{\beta}(z)$ and $I_0(z)$ are the Bessel function and the Bessel function
 of imaginary argument, respectively. Introducing these representations
into (\ref{pro1})
we arrive at the following integral for $\cP_{n}(\mbr,\mbr')$
\beq
\cP_n(\mbr,\mbr')=\frac{1}{(2\pi)^3}\frac{2\alpha^{2}}{s_+-s_-}
\label{pn}
\eeq
$$
\times \int_{-\infty}^\infty dy_1 \int_{-\infty}^\infty dy_2
\int_0^\infty dx_3\,
I_0\left(\sqrt{2x_3\alpha(s_++s_-)}\right)
\int_0^\infty \frac{dx_1}{x_1}\int_0^\infty \frac{dx_2}{x_2}
$$
$$
\times
\left(\frac{e^{-\im y_1}}{1-\im y_1}\right)^n\left(\frac{e^{-\im y_2}}{1-\im y_2}\right)^n
\left(e^{-x_3}\right)^n
\left(x_1e^{-x_1+1}\right)^n\left(x_2e^{-x_2+1}\right)^n
$$
$$
\times \left[ \sqrt{\alpha s_+x_2}J_0(2\sqrt{\alpha s_-x_1})
J_1(2\sqrt{\alpha s_+x_2})- \sqrt{\alpha s_-x_2}J_1(2\sqrt{\alpha
s_-x_2}) J_0(2\sqrt{\alpha s_+x_1}) \right].
$$
By introducing a new five-dimensional variable $X=\{x_1,x_2,x_3,y_1,y_2\}$ and  quantities
$$
\mbox{Q}=
x_1e^{-x_1+1}x_2e^{-x_2+1}e^{-x_3}
\frac{e^{-\im y_1}}{1-\im y_1}\frac{e^{-\im y_2}}{1-\im y_2},
$$
$$
B(X,\alpha,s_{+},s_{-})=
I_0\left(\sqrt{2x_3\alpha(s_++s_-)}\right)
$$
$$
\times
\left[ \sqrt{\alpha s_+x_2}J_0(2\sqrt{\alpha s_-x_1})
J_1(2\sqrt{\alpha s_+x_2})- \sqrt{\alpha s_-x_2}J_1(2\sqrt{\alpha
s_-x_2}) J_0(2\sqrt{\alpha s_+x_1}) \right]
$$
we  rewrite the latter formula in the compact form
\beq
\cP_n(\mbr,\mbr')= \frac{1}{(2\pi)^3}\frac{2\alpha^{2}}{s_+-s_-}
\label{pnc}
\int\limits_{\Omega}\frac{dX}{x_1x_2}
\mbox{Q}^{n}
B(X,\alpha,s_{+},s_{-}).
\eeq
The integration domain $\Omega$ is defined as
\beq
\Omega=\{X:0\le x_{i}<\infty,\  i=1,2,3,\ -\infty < y_k <\infty,\ k=1,2\}.
\eeq
Let us notice that the quantity $|\mbox{Q}|$ is bounded on $\Omega$ with the only  maximum
(such that $|\mbox{Q}|=1$)
at the point
$X_0=\{1,1,0,0,0\}$ , hence everywhere except
$X_0$ the inequality $|\mbox{Q}|<1$ holds true.

Before computing the infinite sum (\ref{P_d}) let us consider the
operator
\beq
\cP_{N_1}^{N_2}=\sum_{n=N_1}^{N_2}\cP_n \nonumber.
\eeq
The kernel of the operator $\cP_{N_1}^{N_2}$ can easily be computed   by using the formula
(\ref{pnc}) for $\cP_n$
and evaluating of the sum of the geometric progression
of $\mbox{Q}^{n}$ terms under the
integral which yield
\beq
\cP_{N_1}^{N_2}(\mbr,\mbr')=P_{N_1}^{N_2}(\mbr,\mbr',\Omega) \equiv
\frac{1}{(2\pi)^3}\frac{2\alpha^{2}}{s_+-s_-}
\label{pn1n2}
\eeq
$$
\times
\int\limits_{\Omega}\frac{dX}{x_1x_2}
\frac{\mbox{Q}^{N_1}(1-\mbox{Q}^{N_2-N_1+1})}{1-\mbox{Q}}
B(X,\alpha,s_{+},s_{-}).
$$

Now we are ready to evaluate the limit of $P_{N_1}^{N_2}(\mbr,\mbr',\Omega)$ as $N_2\to \infty$
keeping $N_1$ finite.
Let us  notice that the integral (\ref{pn1n2}) converges uniformly at any $N_2$ and hence for any
positive $\epsilon$
we can find $\delta>0$  such that
\beq
|P_{N_1}^{N_2}(\mbr,\mbr',\Omega(X_0,\delta))|< \epsilon
\nonumber
\eeq
where $\Omega(X_0,\delta)$ is a neighborhood of the point $X_{0}$ in which the quantity
$\mbox{Q}$ reaches its maximum, i.e.
\beq
\Omega(X_0,\delta)=\{X \in \Omega: |X-X_0|<\delta\}.
\eeq
On the rest of the integration domain $\overline{\Omega}(X_0,\delta)\equiv \Omega \backslash
\Omega(X_0,\delta)$ the
inequality $|\mbox{Q}|<1$ holds true and we can take the limit
\beq
P_{N_1}(\mbr,\mbr',\overline{\Omega}(X_0,\delta))
=\lim_{N_2\to \infty}P_{N_1}^{N_2}(\mbr,\mbr',\overline{\Omega}(X_0,\delta))=
\nonumber
\eeq
\beq
\frac{1}{(2\pi)^3}\frac{2\alpha^{2}}{s_+-s_-}
\int\limits_{\overline{\Omega}(X_0,\delta)}\frac{dX}{x_1x_2}
\frac{\mbox{Q}^{N_1}}{1-\mbox{Q}}
B(X,\alpha,s_{+},s_{-}).
\la{pn1-delta}
\eeq
For the integral (\ref{pn1-delta}) the limit $\delta \to 0$ is permitted, so that due to the
arbitrariness of $\epsilon$
we get
\beq
\cP_{N_1}(\mbr,\mbr')=
\lim_{\delta\to 0} P_{N_1}(\mbr,\mbr',\overline{\Omega}(X_0,\delta))=
\la{pn1i}
\eeq
$$
\frac{1}{(2\pi)^3}\frac{2\alpha^{2}}{s_+-s_-}
\int\limits_{\Omega}\frac{dX}{x_1x_2}
\frac{\mbox{Q}^{N_1}}{1-\mbox{Q}}
B(X,\alpha,s_{+},s_{-}).
$$
Now by setting $N_1=1$ we arrive at the final result for the projection $\cP_d$
\beq
\cP_{d}(\mbr,\mbr')=
\frac{1}{(2\pi)^3}\frac{2\alpha^{2}}{s_+-s_-}
\int\limits_{\Omega}\frac{dX}{x_1x_2}
\frac{\mbox{Q}}{1-\mbox{Q}}
B(X,\alpha,s_{+},s_{-}).
\la{pdi}
\eeq
The formulae (\ref{pn1i}) and (\ref{pdi}) are the main results of this subsection.

\subsection{Some particular cases}
In this subsection we consider a particular case of the  integral representation
(\ref{proj13}) which leads to an analog of the Christoffel-Darboux formula (22.12.1) \cite{abram}
applied to Laguerre polynomials. We also evaluate the asymptotics of the projections kernel
$\cP_n(\mbr,\mbr')$
in the special case when
$n\gg \alpha(r+r')$ with the help of representation (\ref{pro1}).

It is worthwhile to  notice  that the formulae (\ref{pro1},\ref{proj13}) in the particular case when
$n=1,2$
\beq
\cP_1(\mbr,\mbr')=\frac{\alpha^3}{\pi}e^{-\alpha(r+r')},
\nonumber
\eeq
\beq
\cP_2(\mbr,\mbr')=\frac{\alpha^3}{32\pi}e^{-\frac{\alpha}{2}(r+r')}
\left[4-2\alpha(r+r')+\alpha^2rr(1+ \hr\cdot \hr' )\right]
\nonumber
\eeq
give essentially the same results which can be computed directly from the conventional representation
(\ref{P_n},\ref{psi}).
Let us now consider the case of arbitrary $n$. From (\ref{P_n},\ref{psi}) and (\ref{proj13}) we get
\beq
\frac{L_{n-1}(\frac{\alpha s_{+}}{n})L_{n}(\frac{\alpha s_{-}}{n})-
L_{n}(\frac{\alpha s_{+}}{n})L_{n-1}(\frac{\alpha s_{-}}{n})}
{\frac{\alpha}{n}(s_{+}-s_{-})}=
\la{sr1}
\eeq
$$
\frac{1}{n}
\sum_{l=0}^{n-1}\left(\frac{2\alpha r}{n}\frac{2\alpha r'}{n}\right)^{l}
\frac{(n-l-1)!}{(n+l)!}
L^{(2l+1)}_{n-l-1}\left(\frac{2\alpha r}{n}\right)L^{(2l+1)}_{n-l-1}\left(\frac{2\alpha r'}{n}\right)
(2l+1)P_{l}(\hr \cdot \hr').
$$
Taking a particular case $\hr \cdot \hr' =1$, setting $(2\alpha r)/n =x$, $(2\alpha r')/n =y$
and changing the summation variable in such a way
$l=n-m-1$ we get
\beq
\frac{L_{n-1}(x)L_{n}(y)-L_{n}(x)L_{n-1}(y)}
{x-y}=
\label{ourCD}
\eeq
$$
\frac{1}{n}
\sum_{m=0}^{n-1}(xy)^{n-m-1}
\frac{m!(2(n-m)-1)}{(2n-m-1)!}L_{m}^{(2(n-m)-1)}(x)L_{m}^{(2(n-m)-1)}(y).
$$
This formula if compared to the Christoffel-Darboux summation formula for Laguerre polynomials
(22.12.1) \cite{abram}
\beq
\frac{L_{n-1}(x)L_{n}(y)-L_{n}(x)L_{n-1}(y)}
{x-y}=
\label{CD}
\frac{1}{n}
\sum_{m=0}^{n-1}L_{m}(x)L_{m}(y)
\eeq
yields the following interesting identity
\beq
\sum_{m=0}^{n-1}L_{m}(x)L_{m}(y)=
\label{LI}
\eeq
$$
\sum_{m=0}^{n-1}(xy)^{n-m-1}
\frac{m!(2(n-m)-1)}{(2n-m-1)!}L_{m}^{(2(n-m)-1)}(x)L_{m}^{(2(n-m)-1)}(y).
$$

As the last special case we consider the behavior of the projection $\cP_{n}$ for large value of
the principal quantum number $n$. Let us
use again  the representation (\ref{Laguerre}) which we rewrite in the form
$$
L^{(\beta)}_{n-1}(y/n)= \frac{n^n}{n!}n^{\beta+1}y^{-1/2\beta}e^{y/n}F^{\beta}(n,y)
$$
where $F^{\beta}(n,y)$ stands for the integral
$$
F^{\beta}(n,y)=\int_{0}^{\infty}dx\, x^{\beta/2-1}e^{-n(x-\log x)} J_{\beta}(2\sqrt{xy})
$$
and introduce it  into the formula (\ref{pro1}) for Laguerre polynomials.
If $n \gg \alpha (r+r')$ then the only
critical
factor under the respective integrals $F^{\beta}(n,\alpha s_{\pm})$ is $e^{-n(x-\log x)}$ with the
only critical point
$x_0=1$. Evaluating the  the integrals
$F^{\beta}(n,\alpha s_{\pm})$ as $n\to \infty$ by the Laplace method
we get for the projections $\cP_{n}(\mbr,\mbr')$ the following asymptotics
$$
\cP_{n}(\mbr,\mbr')\propto \frac{\alpha^{5/2}}{\pi n^{3}(s_{+}-s_{-})}
[\sqrt{s_{+}}J_0(2\sqrt{\alpha s_{-}})J_{1}(2\sqrt{\alpha s_{+}})-
\sqrt{s_{-}}J_1(2\sqrt{\alpha s_{-}})J_{0}(2\sqrt{\alpha s_{+}})].
$$

\section{Conclusion}
The closed form representations are obtained  for projections onto the $n^2$-dimensional subspace
spanned by bound-state eigenfunctions of the Coulomb Hamiltonian corresponding to the principal quantum
number
$n$ as well as for the projection onto the subspace spanned by all Coulomb bound-states.
These representations can be useful for solving the few body scattering problem in a system of charged
particles for energies above the three body disintegration thresholds.
The asymptotics computed above for the projections $\cP_{n}$ as $n\to \infty$
may lead to  drastic simplifications in calculating
different Coulomb matrix elements between states which are spatially well confined.
The analog of the Christoffel-Darboux summation
formula for Laguerre polynomials which is derived as a particular case of the representations for
the Coulomb projections
can be useful for the theory of classical orthogonal polynomials.

\ack
The work of S. L. Y. was partly
supported by NSF grant Phy-0243740 and INTAS grant No. 03-51-4000. O. M. D. and S. B. L would like to express their gratitude to the Department of Molecular Physics
of Stockholm University for the support  made possible under
the Swedish Institute grant 60886/2005
and Swedish Research Council grant 629-2002-8331.


\end{document}